\documentclass[11pt]{article}
\usepackage{moriond,epsfig}

\def\be{\begin{equation}}
\def\ee{\end{equation}}
\def\bea{\begin{eqnarray}}
\def\eea{\end{eqnarray}}

\def\rc {r_{\rm c}}
\def \ciao {{\sc ciao}} 

\def \nh {$N_{\rm H}$}

\def\xmm{{\it XMM-Newton\/}}
\def\cha{{\it Chandra\/}}
\def\es{erg~s$^{-1}$}

\begin{document}
\vspace*{4cm}
\title{ON THE ENVIRONMENT  OF POWERFUL RADIO GALAXIES AT $z>0.5$}

\author{ E.BELSOLE$^{1,2}$, D. M. WORRALL$^2$, M.J. HARDCASTLE$^3$ }

\address{$^1$Institute of Astronomy, University of Cambridge, Madingley Road, CB3 0HA, Cambridge, UK; $^2$Department of Physics, University of Bristol, Tyndall Avenue, Bristol, BS8 1TL, UK; $^3$ School of Physics, Astronomy and Mathematics, University of Hertfordshire,
 College Lane, Hatfield, Hertfordshire AL10 9AB}

\maketitle\abstracts{
Active galaxies are the most powerful engines in the Universe for
converting gravitational energy into radiation, and radio galaxies
and radio-loud quasars are highly luminous and
can be detected across the Universe. The jets that characterise
them need a medium to propagate into, and thus radio galaxies at high
redshift point to gaseous atmospheres on scales of at least the
radio source diameter, which in many cases can reach hundreds of kpc.
The variation with redshift of X-ray properties of radio-selected
clusters provides an important test of structure formation
theories as, unlike X-ray selection, this selection is not biased
towards the most luminous clusters in the Universe. We present new results from
a sample of 19  luminous radio
galaxies at redshifts between 0.5 and 1. The properties of the
gaseous atmosphere around these sources as mapped by {\it Chandra} and
{\it XMM-Newton} observations are discussed. By combining these with
observations at radio frequency, we will be able to draw conclusions on cluster
size, density, and pressure balance between the radio source and the
environment in which it lies.}

\section{Introduction}
Most powerful radio galaxies lie at high redshift ($z>0.3$), but a few
of them are also found in the nearby universe, the best example of
which is Cygnus A. Sources classified as Fanaroff-Riley type II
(FRII) have double-sided jets which can terminate at distances as far
as 1 Mpc from the centre of the host galaxy. Radio lobes that result
from jet disruption require a gaseous environment in order to exist
and to be confined \cite{begelman86}. This has suggested that powerful
radio galaxies can be used as a means to discover high-redshift
clusters of galaxies (e.g. \cite{lefevre96,fabian01}), which can be
traced by their X-ray emission. Studies based on $ROSAT$ data of radio
galaxies and quasars in the 3CRR catalogue\cite{3crrcat} supported
this hypothesis, and several detection of hot-massive clusters were
claimed \cite{crfab93,dmw94,crfab95,crfab96,mjh98,mjh99,crawford99}. No
spectral confirmation was possible with $ROSAT$ data due to limited
energy sensitivity and count rates. Moreover, the point-like, X-ray
emission from the central radio source together with lobe inverse
Compton (IC) emission, was difficult to separate from any cluster
emission. Hence, detections more often represented upper limits to
cluster emission. \xmm\ and \cha\ have opened a new era of X-ray
astrophysics, and in particular make it possible to carry out studies aimed at searches for clusters
around active galaxies. These studies are fundamental to obtain a
complete, unbiased picture of the number density of clusters and
structure formation and evolution.

{\em Chandra} observations of radio galaxies and quasars at redshift
$z>0.5$ have found relatively few sources in the rich gaseous
environments associated with massive clusters. Given the low number of
counts, for most of them a full spectral analysis was not possible
(e.g. \cite{mjh02,crfab03}), although the detected extended
counts were properly separated from the Point Spread Function
(PSF) of the central point source. For a few objects a spectral
detection did confirm that extended  X-ray emission was associated
with thermal radiation from an intra-cluster medium (3C\,220.1, $kT =
5$ keV \cite{dmw01}; 3C\,294, $kT = 3.5$ keV \cite{fabian03}, although
a large non-thermal contribution to the emission was found to be
possible). The detected atmospheres of other sources are  more typical
of a group or poor cluster (e.g. \cite{crfab03,donahue03}). Although
the very presence of edge-brightened radio lobes points to the
existence of {\em some} gas in order that the lobes should be
confined, a particularly rich environment is not required, and the
{\it Chandra} observations seem to point in this direction. A few more cluster atmospheres were detected spectrally using the higher sensitivity of \xmm\ \cite{bel04}. 

In this paper we present the results of a study based on an unbiased
(although not complete) sample of radio galaxies and quasars in the redshift range $0.5<z<1$ and extracted from the 3CRR catalogue. The sample comprises 19 sources selected as described in Belsole et al. (2006). Here we discuss preliminary results on the environment of the sources.

  Throughout the paper we use a cosmology with $H_{\rm 0}$ = 70 km s$^{-1}$ Mpc$^{-1}$, $\Omega_{\rm m}$ = 0.3, $\Omega_{\Lambda}$ = 0.7. If not otherwise stated, errors are quoted at 1$\sigma$ confidence level.

\section{The sample and data}\label{subsec:prod}

This study is based on the sample described in \cite{bel06}] (hereafter paper I). Sources are drawn from the 3CRR catalogue \cite{3crrcat}, and are in the redshift range $0.5<z<1.0$, with the exception of 3C200 ($z=0.458$). The sample is composed of a similar number of broad-line and narrow-line radio sources. The 3CRR is selected on the basis of low-frequency (178 MHz) radio emission, which has the advantage of not being biased towards beamed radio emission and therefore contains radio galaxies as well as quasars.

\noindent Table 1 lists the main properties of the sources.

\begin{table*}
\caption{The sample and X-ray observations\label{tab:sources}}
\vspace{0.4cm}
\begin{center}
\begin{tabular}{l|cclclcr}
\hline
Source & RA(J2000) & Dec(J2000) & redshift  & scale &type & \nh\\
        & $^{\rm h~m~s}$ &$^{\circ~\prime~\prime\prime}$ & &
        kpc/arcsec & & 10$^{20}$ cm $^{-2}$\\
\hline
3C\,6.1 &  00 16 30.99 & +79 16 50.88  & 0.840 &7.63&  NLRG & 14.80\\  
3C\,184 &  07 39 24.31 & +70 23 10.74  & 0.994 &8.00&  NLRG & 3.45\\
3C\,200 &  08 27 25.44 & +29 18 46.51  & 0.458 &5.82&  LERG & 3.74 \\
3C\,207 &  08 40 47.58 & +13 12 23.37  & 0.684 &7.08&  QSO& 4.12 \\
3C\,220.1& 09 32 39.65 & +79 06 31.53  & 0.610 &6.73&  NLRG & 1.87 \\
3C\,228 &  09 50 10.70 & +14 20 00.07  & 0.552 &6.42&  NLRG & 3.18\\
3C\,254 &  11 14 38.71 & +40 37 20.29  & 0.734 &7.28&  QSO& 1.90 \\
3C\,263 &  11 39 57.03 & +65 47 49.47  & 0.646 &6.90&  QSO& 1.18 \\
3C\,265 &  11 45 28.99 & +31 33 49.43  & 0.811 &7.54&  NLRG & 1.90 \\
3C\,275.1& 12 43 57.67 & +16 22 53.22  & 0.557 &6.40&  QSO& 1.99\\
3C\,280 &  12 56 57.85 & +47 20 20.30  & 0.996 &8.00&  NLRG & 1.13\\
3C\,292 &  13 50 41.95 & +64 29 35.40  & 0.713 &6.90&  NLRG & 2.17 \\
3C\,309.1& 14 59 07.60 & +71 40 19.89  & 0.904 &7.80&  GPS-QSO& 2.30\\
3C\,330 &  16 09 34.71 & +65 56 37.40  & 0.549 &6.41&  NLRG & 2.81 \\
3C\,334 &  16 20 21.85 & +17 36 23.12  & 0.555 &6.38&  QSO & 4.24 \\
3C\,345 &  16 42 58.80 & +39 48 36.85  & 0.594 &6.66&  core-dom QSO&1.13 \\
3C\,380 &  18 29 31.78 & +48 44 46.45  & 0.691 &7.11&  core-dom QSO& 5.67 \\
3C\,427.1& 21 04 06.38 & +76 33 11.59  & 0.572 &6.49&  LERG&  10.90 \\
3C\,454.3& 22 53 57.76 & +16 08 53.72  & 0.859 &7.68&  core-dom QSO & 6.50 \\
\hline
\end{tabular}
\vskip 10pt
\begin{minipage}{15cm}
Galactic column density is from \cite{nh}; NRLG means Narrow Line
Radio Galaxy; LERG means low-excitation radio galaxy. Redshifts and
positions are taken from \cite{3crrcat}.
\end{minipage}
\end{center}
\end{table*}

Preparation of the \cha\ data are described in paper I. However for
this paper we re-processed the data with a later version of \ciao
(3.3.0.1). The \xmm\ data preparation is described elsewhere \cite{bel04}.
 
\section{Results}
Thermal X-ray emission from a cluster-like atmosphere is best detected at energies between 0.5 and 2.5 keV, corresponding to the energy range where X-ray telescopes are most sensitive and where contamination by the central point source is lower (as many cores are absorbed).

\begin{figure}
\hskip -1.5cm
\psfig{figure=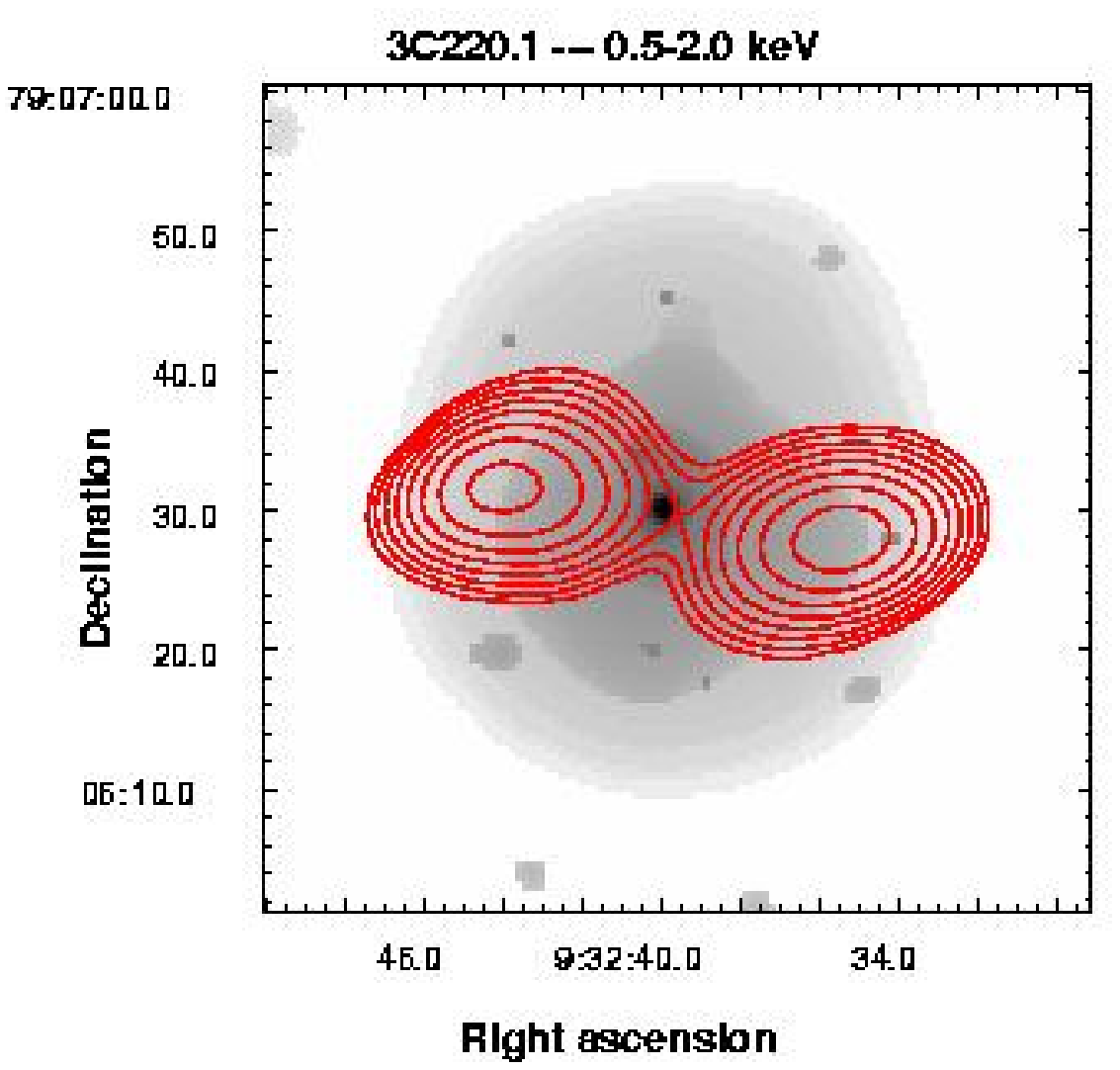,height=3.5in}
\hskip -2.5cm
\psfig{figure=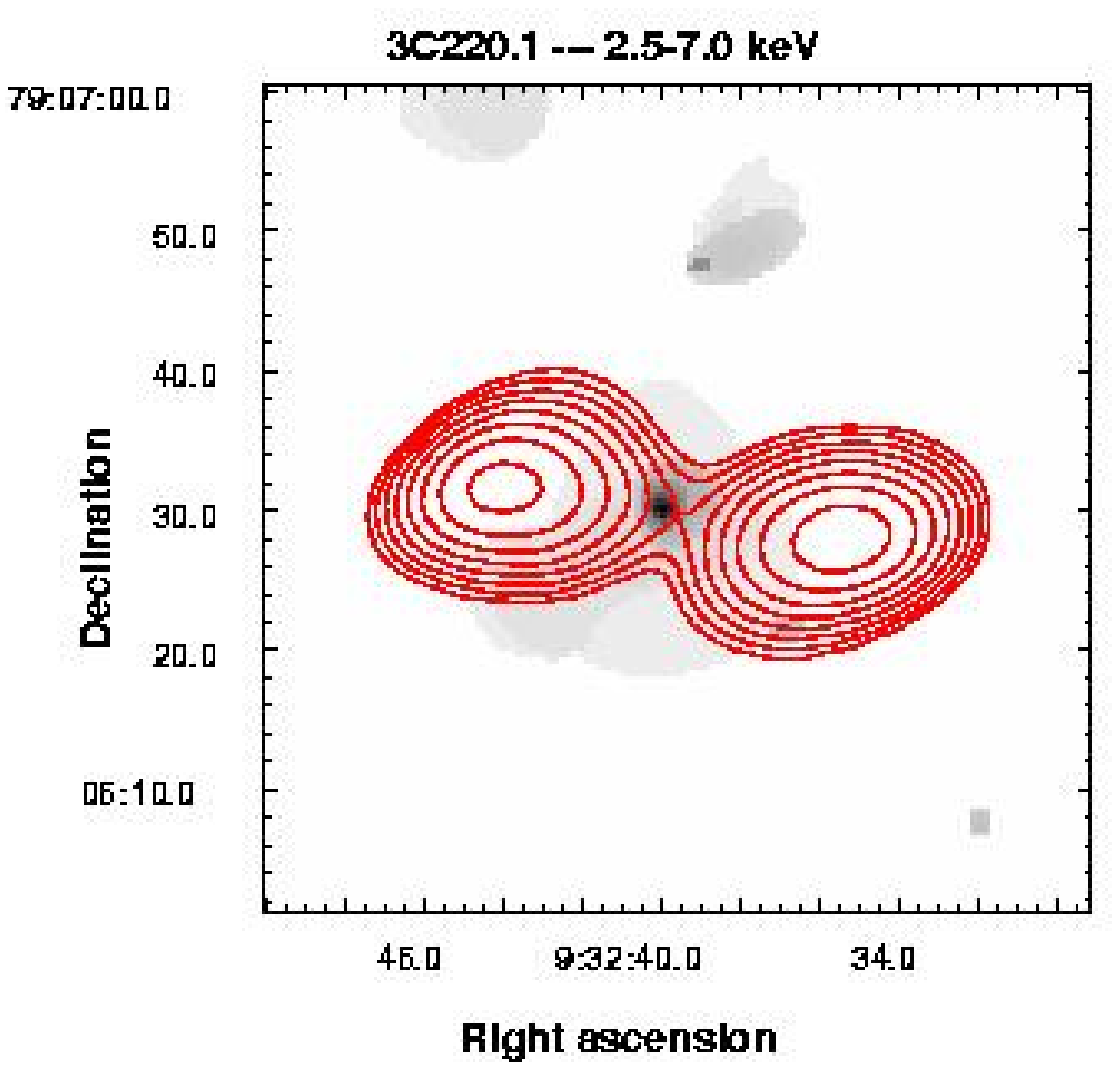,height=3.5in}
\caption{Wavelet reconstructed images in the soft (0.5-2.0 keV, left) and hard (2.5-7.0 keV, right) energy band for the radio galaxy 3C\,220.1. The figure shows that extended emission is visible in the soft band. Contours show the emission at 1.4 GHz and are drawn on a logarithmic scale
\label{fig:3c220.1}}
\end{figure}

We generated images in the 0.5-2.0 keV energy band and in a harder band (2.5-7.0 keV), and we applied a wavelet reconstruction algorithm ($ZHTOOLS$, Vikhlinin private communication) as an efficient way to search for extended emission not associated with the PSF. In Figure \ref{fig:3c220.1} we show 3C\,220.1 as an example of this analysis. Comparison between the images in the soft and hard energy band give an immediate indication of the presence of extended emission.

Extended emission from the 19 sources in the sample was characterised more quantitatively by extracting a radial profile in the 0.5-2.5 keV energy band. Point sources in the field were removed. Each radial profile was modelled with a PSF (generated using $MARX$) appropriate for the position and spectrum of each of the sources. As extended emission may also come from the radio lobes (e.g. \cite{croston05}), we excluded the spatial region coincident with them before extracting the radial profile. When a PSF-only model was not a good representation of the data, a $\beta$-model convolved with the PSF was added to account for the excess emission above the PSF. 

90 per cent of the sources in the sample required the inclusion of  a $\beta$-model to represent the radial profile at soft energies, although a search in the $\beta-\rc$ parameter space (where $\rc$ is the core radius) was not possible for all of them. In these cases, the value of $\beta$ was fixed to 0.6.

We derived the number of counts associated with the $\beta$-model
representing the extended emission out to the radius used for the
background region. This is in the range 35 arcsec to 80 arcsec (which
corresponds roughly to 200-500 kpc by taking an average redshift for
the sources in the sample). We then calculated the X-ray luminosity of
the cluster-like environment. When possible, we carried out spectral
analysis and used the best-fit temperature to calculate the
luminosity. In other cases we adopted a temperature of 3 keV.

\begin{figure}
\psfig{figure=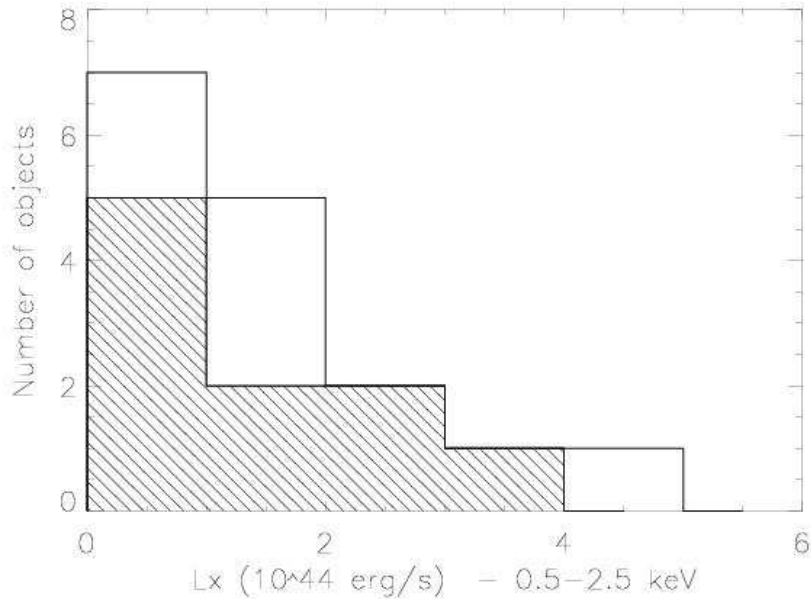,height=3.5in}
\caption{X-ray luminosities of the clusters found around 16 of the objects in the sample. The shaded histogram indicates radio galaxies, the unfilled quasars.
\label{fig:histoLx}}
\end{figure}

In Figure \ref{fig:histoLx} we show the distribution of number of
objects as a function of the X-ray luminosity. The shaded area
corresponds to radio galaxies, the unfilled area to quasars. We
find that 50\% of the radio sources in the sample lie in clusters with
luminosity above 10$^{44}$ \es, and 7 more objects appear to lie in
more moderate luminosity clusters, or groups. Only 1 object shows no
sign of extended emission and is a pure point source at the
sensitivity of this observation. For 2 more objects only upper
limits can be obtained from the radial profile analysis; these are omitted from the histogram above.

We find no obvious correlation between X-ray luminosity and redshift. However, the redshift range in this sample is relatively limited, and a larger sample including sources at lower and higher redshift should be used to be more conclusive.
 
\section{Discussion and conclusions}

This preliminary study, based on a sample of 19 radio galaxies
in the redshift range $0.5<z<1.$ that were selected by their low
frequency radio emission, indicates that a significant fraction of
them (90 per cent) are located at the centre of thermal X-ray emitting
environments. The detected extended emission is unlikely to be
associated with inverse Compton emission from the radio lobes since we
have masked the lobe region from our analysis. 50 per cent  of the
sources in our sample lie in relatively massive clusters, with X-ray
luminosities greater than 10$^{44}$ \es. Although preliminary, these
results suggest that powerful radio galaxies may indeed be an
efficient method to search for luminous (massive) galaxy clusters, and
as a by-product they allow an investigation of a wider population of X-ray structures.

This is particularly important in view of the new generation of X-ray cluster surveys for cosmology: current X-ray cluster catalogues exclude these sources since they are dominated by the emission of the point source, with the result that many objects are only found in Active Galactic Nuclei catalogues and not in cluster catalogues.
They are also likely to be excluded from current and future Sunyaev-Zeldovich (SZ) surveys as the radio emission associated with them is difficult to account for and remove from SZ maps.

This may introduce an important bias in the real number of clusters in the Universe, yielding significant errors (of order 10-20 per cent) on parameters derived from the number density of clusters. Their number can also increase with redshift as shown by Celotti \& Fabian (2004).

To fully understand this issue, \xmm\ and \cha\ observations able to
complete statistical samples of radio galaxies (and the 3CRR sample
looks the most obvious), especially at redshift above 0.1, need to be
carried out now, as future X-ray missions may not have the necessary instrumental characteristics to perform these studies.

\section*{Acknowledgements}
E.B. thanks PPARC for support and MJH thanks the Royal Society
  for a research fellowship. We are grateful to A. Vikhlinin for providing the $ZHTOOLS$  used in the analysis.


\end{document}